\documentclass[12pt]{article}
\usepackage{amssymb,graphicx}
\usepackage{epstopdf}
\usepackage{amsmath,amsfonts}
\usepackage{epsfig}

\def\d{\partial}

\newcommand{\be}{\begin{equation}}
\newcommand{\ee}{\end{equation}}
\newcommand{\bea}{\begin{eqnarray}}
\newcommand{\eea}{\end{eqnarray}}
\newcommand{\bg}{\begin{gather}}
\newcommand{\eg}{\end{gather}}
\newcommand{\bseq}{\begin{subequations}}
\newcommand{\eseq}{\end{subequations}}

\newcommand{\loe}{\lesssim}

\begin{document}
\begin{flushright}
\end{flushright}
\vspace{10pt}
\begin{center}
  {\LARGE \bf   Phantom without UV pathology} \\
\vspace{20pt}
V.A.~Rubakov$^a$\\
\vspace{15pt}
  $^a$\textit{
Institute for Nuclear Research of the Russian Academy of Sciences,\\
60th October Anniversary Prospect, 7a, Moscow, 117312, Russia
  }\\
    \end{center}
    \vspace{5pt}

\begin{abstract}
We present a simple model in which the weak energy condition is violated
for spatially homogeneous, slowly evolving fields. The excitations
about Lorentz-violating background in Minkowski space do not contain ghosts,
tachyons or superluminal modes at spatial momenta ranging from
some low scale $\epsilon$ to the ultraviolet cutoff scale, while 
tachyons and possibly ghosts do exist at ${\bf p}^2 < \epsilon^2$. 
We show that
 in the absence of other matter, slow
roll cosmological regime is possible; in this regime 
$p+\rho <0$, and yet  homogeneity and isotropy are not
completely spoiled (at the expence
of fine-tuning), since for given conformal momentum, the tachyon mode
grows for short enough period of time.

\end{abstract}

\section{Introduction}


There is growing interest in possible exotic
gravitational phenomena at
cosmological distances, triggered in part by the observation of
the accelerated expansion of the Universe and in part by theoretical
considerations. Approaches
in this direction include the the introduction of new
light fields and the modification of gravity at
ultra-large scales. Among the former, 
phantom energy with equation of state $p+\rho<0$,
thus violating weak energy condition,
is certainly of   interest both from purely theoretical viewpoint
and possibly also for cosmology~\cite{phantomcosmology}. If phantom 
were to dominate the late-time cosmological evolution, a potentially
observable feature would be the accelerated acceleration of the 
Universe\footnote{This effect may be mimicked~\cite{mimic} 
in scalar-tensor theories
of gravity with matter obeying
$p+\rho>0$, see, however Ref.~\cite{however}.};
if phantom were to drive inflation, the spectrum of relic gravitational waves
would be blue, which would be  potentially observable as well.
However, most field theories violating weak energy condition, 
constructed to date,
are pathological
in the ultraviolet domain (UV;
hereafter by  UV  we mean  sufficiently high
energies, but still below a UV cutoff of an effective
field theory): 
they have either ghosts or tachyons or superluminally propagating 
modes, or  combinations  thereof. 
Furthermore, there are general arguments implying that
in four dimensions, 
any\footnote{ A possible exception has to do with
 (spontaneous) violation
of 3d rotational invariance~\cite{260}.}
such theory has UV 
instabilities~\cite{Vikman,Hsu,260}. Yet, purely phenomenologial
analysis~\cite{Hu} reveals that this situation may not be
completely generic. Also, the brane-world DGP model~\cite{DGP} 
which may be well-behaved in UV~\cite{NR},
allows for a period of accelerated cosmological expansion
with effective $p+\rho < 0$~\cite{Sahni,Lue}. 
So, it makes sense to try to construct
an example of a four-dimensional
field-theoretic phantom with non-pathological
UV behavior.

In view of the arguments of Ref.~\cite{260},
it is rather unlikely that there exist field theories with
$p+\rho < 0$ and no problematic features 
at all. As a modest approach,
one may begin with a field theory in Minkowski space,
which is consistent at  
energy scales from zero all the way 
up to the UV cutoff scale ${\cal M}$ of an effective theory. One
then deforms this theory in IR in such a way that its high
energy behavior  remains 
healthy, while the weak energy condition is
violated for spatially homogeneous configurations, and 
pathological states appear below a certain low scale
$\epsilon$ only\footnote{In theories with ghosts and unbroken
Lorentz symmetry this would not be
acceptable anyway, since instabilities would occur at all
spatial momenta and frequencies. Once Lorentz-invariance is broken,
this observation no longer applies, and a theory with ghosts
may be viable even
for rather high IR scale~\cite{Cline}.}. When pursuing this approach,
one has to worry about possible superluminal modes which may emerge
at high momenta after the theory is deformed in IR: this effect,
even of small magnitude, would signal an inconsistency
of the whole theory~\cite{AHetal}.

In cosmological setting, a theory of this sort may be acceptable
provided that the energy scale $\epsilon$ is below,
or at least sufficiently close to the Hubble scale. One envisages
that the latter property needs fine tuning over and
beyond other adjustments of
parameters required to make the theory consistent with observations.

One way to implement this approach is to begin with a more 
or less conventional two-derivative theory\footnote{Another 
interesting possibility
may emerge in a ghost condensate model~\cite{ghostcondensate}
with negatively tilted potential~\cite{senatore}. In the
expanding Universe,
this model indeed has $p+\rho < 0$. A potential problem with this
model is wrong sign of the
quadratic gradient term in the action for 
perturbations~\cite{ghostinflation}; it remains to be understood
how dangerous this feature is.}
  with healthy behavior
below the scale ${\cal M}$, at least in certain backgrounds,
and then add one-derivative terms to the Lagrangian, the latter suppressed
by the small parameter $\epsilon$. Such a construction is
difficult to realize in  scalar theories, so one is naturally lead to 
consider theories involving vector field(s). By now it is understood
that  UV problems
 inherent in vector theories without gauge invariance may be 
avoided~\cite{Ben}, and that one-derivative terms indeed give 
rise to interesting IR dynamics~\cite{MLVR}.

In this paper we construct a simple model along these lines.
The model is described in Section~\ref{sec:Model}.
The background is assumed to break 
Lorentz-invariance, but leave three-dimensional
rotational invariance unbroken. We will check in 
Section~\ref{sec:Flat} that in flat space-time, there are
no ghosts, tachyons or superluminal modes at
three-momenta above the critical momentum $p_c \sim \epsilon$.
At $p<p_c$ there are tachyons and possibly ghosts.

Turning to the cosmological evolution in Section~\ref{FRW}, 
we will show that, 
in the absence of other matter, a slow roll regime is
possible  with appropriate choice
of the potential, once 
the homogeneous
fields filling the Universe take appropriate values.
In this regime, the equation 
of state 
corresponds to $p+\rho <0$, with 
$|p+\rho| \ll \rho$, and, indeed, the fields slowly roll {\it up}
the potenial. In the model we consider, the slow roll
regime occurs at $H \ll \epsilon$, so that the tachyonic mode
is sub-horizon for some period of time.

We will then study inhomogeneous perturbations of the fields
about slowly-rolling background (Section~\ref{sec:Pert}). 
We will show that for given conformal momentum $k$, the  dangerous 
tachyonic mode grows
for finite  period of time,
until it becomes super-horizon. 
Super-horizon modes do not grow; some of them, including
would-be tachyon, freeze in (like in the
case of the minimal scalar field in accelerating Universe), 
some decay. This property
is not entirely trivial for the would-be tachyonic mode,
and has to do with the fact that this mode is gapless in 
Minkowski background.
With extra fine-tuning, the period of growth of the tachyonic mode
can be made
sufficiently small, so that the model can be made viable.
We conclude in Section~\ref{sec:Conclusions}.

\section{Model}
\label{sec:Model}

The model we wish to present has two-derivative kinetic terms
similar to those in Ref.~\cite{Ben}, one-derivative term
reminiscent of that in Ref.~\cite{MLVR}, and also a potential term. 
It has one vector field $B_\mu$ and one
scalar field $\Phi$. The Lagrangian is\footnote{Space-time signature
is $(-,+,+,+)$.}
\be 
   L= L^{(2)} + L^{(1)} + L^{(0)}
\label{modeltot}
\ee
where superscripts indicate the number of derivatives, and we take
\bea
L^{(2)} &=& 
 - \frac{1}{2} \alpha (\xi)
           g^{\nu \lambda } D_\mu B_\nu  D^\mu B_\lambda 
           + 
\frac{1}{2}\beta(\xi)
D_\mu B_\nu   D^\mu
      B_\lambda \cdot \frac{B^\nu B^\lambda}{{\cal M}^2}
    +\frac{1}{2} \partial_\mu \Phi \partial^\mu \Phi
\label{Lmodel-2} \\
L^{(1)} &=& \epsilon \partial_\mu \Phi B^\mu
 \\
L^{(0)} &=& -V(B, \Phi)
\eea
where
\[
   \xi = \frac{B_\mu B^\mu}{{\cal M}^2}
\]
Here $\alpha$ and $\beta$ are functions roughly of order 1,
and $\epsilon$ is a free positive 
parameter -- the IR scale;
${\cal M}$ can be viewed as the UV cutoff scale of the effective theory,
so that ${\cal M} \gg \epsilon$. We will be interested in Lorentz-violating
backgrounds 
with $\xi \neq 0$.

The Lagrangian is Lorentz-invariant provided that the potential $V$
depends on a combination $X= \sqrt{B_\mu B^\mu}$. For simplicity we
will mostly consider potentials of the form
\be
   V=U(X) + W(\phi)
\label{separable}
\ee
We begin with Minkowski space, and consider 
static homogeneous background 
\be
B_0 = X =\mbox{const} \; , \;\;\; B_i=0 \; , \;\;\;\; 
\Phi=\phi = \mbox{const}
\label{bck1}
\ee
where $B_0$ and $B_i$ are time and space components of $B_\mu$.
We will be interested in 
small inhomogeneous perturbations 
\be
B_0 = X + b_0 \; , \;\;\;
B_i = b_i\;,\;\;\; \Phi=\phi + \varphi
\label{defpert}
\ee
Their mass terms  are
\be
-\frac{m_0^2}{2} b_0^2 - \frac{m_1^2}{2} b_i^2 -
\frac{m_\varphi^2}{2} \varphi^2
\label{massterms}
\ee
with
\bea
  m^2_{0} &=&  U_{XX}  \;, 
\label{mb0}\\
  m_{1}^2 &=& - \frac{1}{X} U_X\;, 
\label{mbi}\\
  m_\varphi^2 &=&  W_{\phi \phi}
\label{mvarphi}
\eea
Here we introduced convenient notation for derivatives,
\[
         \frac{\partial U}{\partial X} 
\equiv  U_X\;, \;\;\;\;
 \frac{\partial^2 U}{\partial X^2} \equiv U_{XX}\;, \;\;\; \mbox{etc.}
\]


Let us make a point concerning the energy-momentum tensor. For a field
configuration with $B_i=0$, Noether's energy-momentum tensor
is
\[
  T_{\mu\nu} = \kappa \d_\mu B_0 \d_\nu B_0 + 
               \d_\mu \Phi \d_\nu \Phi 
               + \delta_{\mu 0} ~\epsilon ~\d_\nu \Phi B_0
               - \eta_{\mu \nu} L
\]
where
\be
    \kappa (X) = \frac{X^2}{{\cal M}^2} \beta (X) - \alpha (X)
\label{kappadef1}
\ee
In what follows we assume 
\[
\kappa > 0
\]
If $B_0$ and $\Phi$ are spatially homogeneous,
$B_0 = X(t)$, $\Phi= \phi(t)$, 
energy density and pressure are
\bea
  \rho  &=& \frac{\kappa}{2} \dot{X}^2 + \frac{1}{2} \dot{\phi}^2
           + V \nonumber \\
    p   &=& 
 \frac{\kappa}{2} \dot{X}^2 + \frac{1}{2} \dot{\phi}^2
+ \epsilon \dot{\phi} X
          -  V 
\nonumber
\eea
The relation $p+\rho <0$ is satisfied provided the time deivatives
of the fields are small, and
\be
    \dot{\phi} X  < 0
\label{dotphiX}
\ee
In the cosmological context, time derivatives will indeed be small
if the fields slowly roll along the potential; we will see later 
how this requirement, as well as
the relation (\ref{dotphiX}), can be satisfied.

\section{Minkowski spectrum}
\label{sec:Flat}

\subsection{Massless case}

Let us switch gravity off and
consider the spectrum of this model in Minkowski space and 
in background (\ref{bck1}). 

To begin with, we 
neglect the mass terms and write the quadratic Lagrangian for 
perturbations,
\be
L_{b_i,b_0, \varphi} = \frac{\alpha}{2} \d_\mu b_i \d^\mu b_i
                + \frac{\kappa}{2} \d_\mu b_0 \d^\mu b_0
                + \frac{1}{2} \d_\mu \varphi \d^\mu \varphi
                + \epsilon \d_0 \varphi b_0 - 
                  \epsilon \d_i \varphi b_i
\label{Lquadr}
\ee
From the Lagrangian (\ref{Lquadr}) it is straigtforward to find
the spectrum. The transverse components of $B_i$
decouple and have the standard massless dispersion relation 
$\omega^2 = p^2$. Among the three remaining modes --- linear
combinations of $b_i = (p_i/p) b_L$, $b_0$ and $\varphi$,
there exists one mode with $\varphi=0$, $b_0 = b_L$
and the same disperison relation, 
$\omega^2 = p^2$, and two modes with dispersion relations
\be
  \omega^2 = p^2 + \frac{\epsilon^2}{2\kappa}
  \pm \sqrt{\left( \frac{1}{\alpha} + \frac{1}{\kappa}\right)
  \epsilon^2 p^2 + \frac{\epsilon^4}{4\kappa^2}}
\label{dispgen}
\ee
These modes are not ghosts at any $p$.
Let us consider their  properties in two
different ranges of momenta.

We begin with high momenta, 
$p^2 \gg (\alpha^{-1} + \kappa^{-1}) \epsilon^2$.
Expanding  in inverse momentum, we obtain
\be
   \omega = p \pm \frac{1}{2} \epsilon
\sqrt{\frac{1}{\alpha} + \frac{1}{\kappa}}
+ \frac{1}{8} \frac{\epsilon^2}{p}
\left( \frac{1}{\kappa} - \frac{1}{\alpha}\right)
\label{omegahigh}
\ee
We see that the modes have group velocity
$\d \omega /\d p$ smaller than the speed of light provided that
\be
    \alpha > \kappa
\label{akineq}
\ee
We concentrate on this case in what follows.

Let us turn to low momenta. There is a critical momentum, 
\be
p_c^2=\frac{\epsilon^2}{\alpha}
\label{maxmom}
\ee
Above this momentum, no tachyonic modes exist. At $p^2<p^2_c$
one of the
modes (that with the negative sign in (\ref{dispgen})) is tachyonic.
The highest possible tachyonic $|\omega^2|$ is at
\[
 p^2 = \frac{\epsilon^2}{4\alpha}~\frac{2\alpha + \kappa}{\alpha+\kappa}
\]
and at this momentum one has
\be
  \omega^2_{min} = - \frac{\kappa}{4(\alpha + \kappa)} 
\frac{\epsilon^2}{\alpha}
\label{omegamin}
\ee
It is worth noting that this value is much smaller than the
maximum tachyonic momentum squared (\ref{maxmom}) provided that
\be
    \alpha \gg \kappa \; ,
\label{aka}
\ee
which of course requires a sort of fine-tuning.
%

Thus,  the massless model in Minkowski background
does not have superluminal,
ghost or tachyonic modes at high momenta, and have 
tachyons at low momenta.

\subsection{Non-zero masses}
\label{sec:Minkowski}

Let us now turn on masses of the perturbations,
i.e., add the terms (\ref{massterms}) to the Lagrangian
(\ref{Lquadr}).
In what follows we will be interested in
the case
\be
    m_{\varphi}^2=0 \; , \;\;\;
m_0^2 \sim m_1^2 \; , \;\;\;
\epsilon^2 \gg m_0^2, m_1^2
\label{massrel}
\ee
Furthermore, we assume for simplicity of the analysis that
the relation (\ref{aka}) holds.

The transverse modes $b_i$ again decouple; they have the dispersion
relation
\be
   \omega_{transv}^2 = p^2 + \frac{m_1^2}{\alpha}
\label{otrans}
\ee
Equations for $b_0$, $\varphi$ and $b_L$ are
\bea
    \left(\omega^2 - p^2 - \frac{m_0^2}{\kappa} \right) b_0
    + i \frac{\epsilon}{\kappa} \omega \varphi &=& 0
\nonumber \\
    \left(\omega^2 - p^2 - \frac{m_1^2}{\alpha} \right) b_L
    - i \frac{\epsilon}{\alpha} p \varphi &=& 0
\nonumber \\
\left(\omega^2 - p^2 \right) \varphi
    - i \epsilon \omega b_0 + i \epsilon p b_L &=& 0
\label{lineq}
\eea
The analysis of high momentum modes goes through almost
as before. Expanding $\omega$ in inverse momentum,
one obtains under our assumptions that the two modes
(\ref{omegahigh}) remain the same (modulo small corrections), 
while the third mode behaves
as
\be
   \omega = p + \frac{m_0^2 + m_1^2}{\alpha + \kappa}
\frac{1}{2p}
\label{3dhigh}
\ee
For positive $m_0^2$ and $m_1^2$ this mode is also subluminal.
All modes have healthy kinetic terms at high momenta, so there are no
ghosts in UV.

Let us consider now the low-momentum modes, and assume 
$m_0^2 > 0$, $m_1^2 >0$ (we will comment on the case 
$m_0^2 < 0$, $m_1^2 >0$ in the end of this section).
It is straightforward to see that the tachyonic mode again
exists at $p^2 < p_c^2$, and the lowest tachyonic $\omega^2$ 
is still given by (\ref{omegamin}). It is instructive to study
the
dispersion relations at very low momenta, $p^2 \ll m_1^2 /\alpha$.
By inspecting eqs.~(\ref{lineq}), one finds that one mode contains
mostly  the field $b_L$
and has 
\be
   \omega_{b_L}^2 = \frac{m_1^2}{\alpha}
\label{obi}
\ee
There are two more modes:  one non-tachyonic mode with
\be
    \omega_{normal}^2 = \frac{\epsilon^2}{\kappa}
\label{oeps}
\ee
and one tachyonic mode with
\be
   \omega_{tachyonic}^2 = - \frac{m_0^2}{m_1^2} p^2
\label{omegatach}
\ee
Neither of the modes with positive $\omega^2$ is a ghost.
It is important that  $\omega^2$ of the tachyonic mode
vanishes as $p^2 \to 0$; in fact, even though eq.~(\ref{omegatach})
as it stands
is valid for $\epsilon^2 \gg m_0^2, m_1^2$ only, the propery that 
the tachyonic mode is gapless, 
$\omega_{tachyonic}^2 \propto -p^2$ as $p^2 \to 0$,
is of general validity.

To end up this section, 
let us comment on the case 
\[
m_0^2 <0 \; , \;\;\; m_1^2 >0
\]
still assuming that (\ref{massrel}) holds for absolute values.
It follows from (\ref{omegahigh}) and (\ref{3dhigh})
that in this case all modes are still subluminal at high momenta, 
provided that 
\[
         |m_0^2| < m_1^2
\]
The tachyonic mode again appears at $p^2=p_c^2$. The special
property of this case is that at even lower momenta, 
$p^2 < |m_{0}^2|/ \kappa$, this mode  again becomes 
non-tachyonic, and  there appears a negative energy state ---
ghost in the spectrum.
Consider, e.g., the special case
\be
    m_0^2 = - m_1^2 < 0
\label{specialmasses}
\ee
In this case,
the tachyon exists in a finite interval of momenta\footnote{The tachyon
mode can be avoided completely at the expence of fine-tuning
$\frac{m_1^2}{\kappa} =\frac{\epsilon^2-m_1^2}{\alpha}$. In that case
the only potentially 
dangerous feature of the model is a ghost in IR, i.e.,
at $p<p_c$. This ghost is phenomenologically acceptable
 even for relatively 
large $p_c$, well exceeding the Hubble parameter~\cite{Cline}.},
$\frac{m_1^2}{\kappa} <p^2<\frac{\epsilon^2-m_1^2}{\alpha}$.
There is a mode with massless dispersion relation at all momenta,
\be
\omega^2 = p^2
\label{specrel}
\ee
together with the  modes (\ref{omegahigh})
which are subluminal  at high momenta.
At high $p^2$ the mode (\ref{specrel}) has positive energy, while 
precisely at that value of momentum for which the tachyon disappears,
$p^2 = m_{1}^2/ \kappa$, this mode becomes a ghost.
It remains to be the ghost at all momenta below $m_{1}^2/ \kappa$.

\section{Cosmological evolution}
\label{FRW}

\subsection{Field equations}

In the case of spatially homogeneous fields with $B_i=0$
in spatially flat FRW metric
\[
  ds^2 = N^2 (t) dt^2 - a^2(t) d{\bf x}^2
\]
the Lagrangian (\ref{modeltot}) reads
\be
\sqrt{g} L = \frac{\kappa}{2} \frac{a^3}{N}\dot{X}^2
- \frac{3\alpha}{2} \frac{\dot{a}^2 a}{N} X^2 + \frac{1}{2}
\frac{a^3}{N}\dot{\phi}^2
+ \epsilon a^3 \dot{\phi} X - a^3N V(X,\phi)
\label{Lhom}
\ee
where $X= N^{-1}B_0$.
It is worth noting that, up to
an $X$-dependent factor, the second term here is proportional to
the gravitational Lagrangian, $\sqrt{-g}R \propto (\dot{a}^2 a)/N$
modulo total derivative.
Therefore, the effective ``cosmological'' Planck mass, entering the
Friedmann equation, is modified as follows (cf. Ref.~\cite{effM}),
\be
    M_{Pl}^2 \to M_{Pl}^2 + \frac{\alpha}{4\pi} X^2
\label{effMpl}
\ee
We will have to make sure that this change is small\footnote{Another
possibility is
that
$X$ varies in time sufficiently slowly. We will not need to resort to 
this option.}.

From the Lagrangian (\ref{Lhom}) one obtains the energy density and 
pressure,
\bea
\rho &=& -\frac{1}{a^3}\left[ 
\frac{\delta S_{X,\phi}}{\delta N} \right]_{N=1} =
\frac{\kappa}{2} \dot{X}^2 -\frac{3\alpha}{2}H^2 X^2
+ \frac{1}{2} \dot{\phi}^2 + V  
\nonumber \\
p&=& \frac{1}{3a^2}\left[
\frac{\delta S_{X,\phi}}{\delta a} \right]_{N=1} =
\frac{\kappa}{2} \dot{X}^2 +\frac{3\alpha}{2}H^2 X^2
+ \alpha \dot{H} X^2 + 2\alpha H \dot{X} X
+ \frac{1}{2} \dot{\phi}^2 + \epsilon \dot{\phi} X
- V  
\nonumber
\eea
where $H=\dot{a}/a$ is the Hubble parameter, and we set $N=1$ for
the rest of this section.
The quantity of interest is
\be
\rho + p =  \epsilon \dot{\phi} X + \alpha \dot{H} X^2 + 
2\alpha H \dot{X} X + \kappa \dot{X}^2 + \dot{\phi}^2
\label{p+rho}
\ee
The most important for our purposes is the first term here.

Equations of motion for homogeneous fields are
\bea
- \kappa \left(\ddot{X}  + 3H\dot{X} \right) -
\frac{1}{2}\kappa_X \dot{X}^2
-\frac{3}{2}\alpha_X H^2 X^2
-3\alpha H^2 X  + \epsilon \dot{\phi} &=&  V_ X
\label{eqX} \\
- (\ddot{\phi} + 3H \dot{\phi}) - \epsilon (\dot{X} + 3HX)
&=&  V_\phi
\label{eqphi}
\eea
We will assume in what follows that $\alpha(X)$ and 
$\kappa (X)$ are not rapidly changing functions,
so that
\be
  |\alpha_X| \loe \frac{\alpha}{X}
\label{8*}
\ee
\be
  |\kappa_X| \loe \frac{\kappa}{X}
\label{8+}
\ee
We are going now to discuss slow roll regime.

\subsection{Slow roll}
Besides the usual requirement that the fields and the Hubble parameter
evolve slowly, 
we define the particular slow roll regime of interest for our purposes 
as the regime at which
terms
proportional to $\epsilon$
dominate in the left hand sides of the field equations
(\ref{eqX}), (\ref{eqphi}).
The slow roll conditions
are therefore not exactly the same here as in the case
of 
inflation driven by scalar inflaton 
(or acceleration driven by scalar quintessence). 
One point is that there are terms without time
derivatives in the field equation
(\ref{eqX}). These terms are undesirable, so one of our 
conditions is
\be
   \alpha H^2 X \ll \epsilon \dot{\phi}
\label{roll6}
\ee
In view of (\ref{8*}) this condition implies that both
non-derivative terms in eq.~(\ref{eqX}) are small.
Hereafter when writing inequalities, we will always mean
absolute values of the quantities on the left and right hand sides.

Another non-trivial slow-roll condition that ensures that
$\dot{\phi}$-term is subdominant in eq.~(\ref{eqphi}) is
\be
     \dot{\phi} \ll \epsilon X
\label{roll5}
\ee
Other conditions are fairly obvious:
\bea
     \dot{X} &\ll& HX
\label{roll2} \\
     \ddot{\phi} &\ll& H \dot{\phi}
\label{roll3} \\
 \dot{H} &\ll& H^2
\label{roll1}
\eea
Note that because of (\ref{8*}) and (\ref{8+}),
$\epsilon \dot{\phi}$-term indeed dominates over all other terms
 in the left hand side of 
eq.~(\ref{eqX}), provided that
(\ref{roll6}) and (\ref{roll2}) are satisfied.
Finally, the potential term dominates the cosmological evolution
provided that
\be
   V \gg \epsilon \dot{\phi} X
\label{rolladd}
\ee
Once these slow-roll conditions are satisfied,   
the field equations take simple form,
\bea
   \epsilon \dot{\phi} =  V_X
\label{eqroll1}
\\
   3H\epsilon X =  -  V_\phi
\label{eqroll2}
\eea
In this slow roll regime, the first term in 
eq.~(\ref{p+rho}) dominates, and 
\[
  \rho + p = X V_X
\]
which is negative for
\be
    V_X <0
\label{negVX}
\ee
(without loss of generality we assume that $X>0$). 

Equation~(\ref{eqroll2}) is algebraic, so the whole system
of the slow-roll equations may appear unusual. To understand it better,
it is instructive to consider the potentials of the
form (\ref{separable}) and view
eq.~(\ref{eqroll1}) as an equation for $X$, whose solution is
\[
   X = F(\dot{\phi})
\]
Keeping only the terms proportional to $\epsilon$ in eq.~(\ref{eqphi}),
we write  eq.~(\ref{eqphi}) as follows
\be
   \epsilon [F^\prime(\dot{\phi}) \ddot{\phi} + 3HF(\dot{\phi})]
= - W_\phi
\label{explain}
\ee
This equation can be understood as the field equation in a
scalar field theory with the action
\be
\int~d^4x~\sqrt{g} \left[ K(\dot{\phi}) - W(\phi) \right]
\label{explain2}
\ee
where 
\[
K(z)= \epsilon \int~F(z)~d z
\] 
Truncated equation (\ref{eqroll2}) is then
the slow-roll equation for this theory. As an example,
for
\be
   U(X) = -\frac{M^2}{2} X^2
\label{UM}
\ee
one has from eq.~(\ref{eqroll1}) 
\[
  X = - \frac{\epsilon}{M^2} \dot{\phi}
\]
and equation (\ref{explain}) reads
\[
\frac{\epsilon^2}{M^2} (\ddot \phi +3H\dot \phi) = + W_\phi  
\]
which is the usual scalar field equation, but in upside-down
potential. In other words, the kinetic term in the effective
Lagrangian (\ref{explain2}) has negative sign. This explains 
why $\rho+p < 0$ in the slow roll regime. 

Let us come back to
the general case. 
Making use of eqs.~(\ref{eqroll1}) and (\ref{eqroll2}) one can rewrite 
the slow-roll conditions in terms of the potential $V(X,\phi)$ and 
its derivatives, or, in the special case (\ref{separable}),
in terms of $U(X)$, $W(\phi)$ and their derivatives. 
As an example, assuming that (\ref{roll1}) is 
satisfied, one finds from (\ref{roll2})
\be
\frac{1}{H\epsilon}  W_{\phi \phi} U_{X} \ll W_\phi
\label{Roll2}
\ee
For potentials obeying 
\be
U_X \sim U/X\; , \;\;\;
W_{\phi} \sim W/\phi\;, \;\;\; W_{\phi \phi} \sim W/ \phi^2
\label{powerlaw}
\ee
(e.g., for power-law potentials) one makes use of
eq.~(\ref{eqroll1}) again, and finds from the latter
inequality that
\[ 
     W \gg U
\]
Thus, the potential $W(\phi)$ dominates in the Friedmann
equation, so that
\[
 H^2 = \frac{8\pi}{3M_{Pl}^2} W
\]
It is now clear that the energy density indeed increases
in time in our slow roll regime. Assuming without loss
of generality that $X>0$, one finds from (\ref{eqroll2})
that $W_\phi<0$, while from (\ref{eqroll1}) and (\ref{negVX})
it follows that $\dot{\phi} <0$. This means that the field $\phi$
climbs the potential up; the value of $W(\phi)$, and hence the
energy, increases.

The condition (\ref{roll3}) gives
\bea
\frac{1}{(\epsilon H)^2} U_{XX} W_{\phi \phi} &\ll& 1
\label{Roll3} \\
\frac{\dot{H}}{H^2} \frac{1}{H\epsilon} U_{XX} W_\phi &\ll& U_X
\eea
For potentials obeying (\ref{powerlaw}), one makes use of
eq.~(\ref{eqroll2}) and finds that the 
second relation here is equivalent to (\ref{roll1}).
The relations (\ref{Roll2}) and (\ref{Roll3}) suggest that the
potentials are sufficiently flat.

The condition (\ref{roll6}) gives the relation of somewhat different 
sort,
\be
       W_\phi \ll \frac{\epsilon}{\alpha H} U_X 
\label{co1}
\ee
Similarly, one finds from (\ref{roll5})
\be 
       U_X \ll \frac{\epsilon}{H} W_\phi
\label{co2}
\ee
These can be satisfied simultaneously only for
\be
    \frac{\epsilon^2}{\alpha H^2} \gg 1
\label{largeepsilon}
\ee
Finally, one finds from (\ref{roll1})
\[
    \frac{M_{Pl}}{\epsilon W^{3/2}} W_\phi U_X \ll 1
\]

To see that all above conditions can indeed be satisfied,
let us begin with the example (\ref{UM}). In this case the evolution
of the  field $\phi$ and the scale factor corresponds to the action
(\ref{explain2}) with 
\[
  K = - \frac{\epsilon^2}{2M^2} \dot{\phi}^2
\]
Hence, the slow roll conditions (\ref{roll2}), (\ref{roll3}) and
(\ref{roll1}), written in terms of the field 
\[
   \sigma = \frac{\epsilon}{M} \phi
\]
are the same as in inflationary theory; for power-law potentials
$W$ they give $\sigma \gg M_{Pl}$, i.e.
\be
   \phi \gg \frac{M M_{Pl}}{\epsilon}
\label{infl}
\ee
The validity of (\ref{roll6}) and (\ref{roll5}) is not guaranteed,
however, so these conditions are to be imposed in addition to
(\ref{infl}). One obtains from (\ref{roll5})
\be
    M^2 \ll \epsilon^2
\label{addM1}
\ee
while (\ref{roll6}) gives
\be
    \frac{M^2}{\alpha} \gg H^2
\label{addM2}
\ee
The two relations (\ref{infl}) and (\ref{addM2}) 
are generally 
satisfied simultaneously only in a finite interval of the values of
$\phi$; for example, for $W=(1/2)\mu^2\phi^2$ this interval is
\[
     \frac{M M_{Pl}}{\epsilon}  \ll |\phi| \ll 
\frac{M M_{Pl}}{\sqrt{\alpha} \mu}
\]
(note that because of (\ref{eqroll2}), slow-roll dynamics
occurs at $\phi <0$ for positive $X$)
which requires 
\[
    \mu^2 \ll \frac{\epsilon^2}{\alpha}
\]
To end up with this example, we note that (\ref{rolladd}) is equivalent
to one of the standard slow-roll conditions
\be
   \dot{\sigma}^2 \ll W
\label{dotchi}
\ee
and that one indeed has
\[
  \alpha X^2 \ll M_{Pl}^2
\]
(see discussion after eq.~(\ref{effMpl})), due to the relations
(\ref{addM2}) and (\ref{dotchi}). Thus, our first example indeed
provides the case in which all our requirements are satisfied.

The above example has the property that $U_{XX} <0$,
so that the mass (\ref{mb0}) is negative. In fact, in this example
the masses obey (\ref{specialmasses}), so there are no superluminal
modes and/or ghosts at high momenta. The 
existence of the ghost (\ref{specrel}) at low momenta is in accord with
the negative sign of $K(\dot{\phi})$ in (\ref{explain2}).
If one adds extra terms 
to the potential (\ref{UM}), the mode (\ref{specrel}) becomes
either subluminal or superluminal at high spatial momenta.

An example
which does not necessarily have $U_{XX} <0$ is determined, e.g., 
by potentials
\bea
   U(X) &=& \zeta^2 X_0^4 u\left(\frac{X}{X_0} \right)
\nonumber \\
 W(\phi) &=& \tau^2 \phi^4
\label{12*}
\eea
where $X_0$ is some typical value of the field $X$,
$u$ is a smooth function of order 1 whose  derivatives are also
of order 1 and
$\zeta$ and $\tau$ are coupling constants (in this example we set 
$\alpha=1$ by redefining the fields). The quartic potential
(\ref{12*}) is chosen for concreteness only. 
Equation (\ref{eqroll2}) tells that 
the slow roll occurs at $X\sim X_0$ if
\[
    \phi \sim \frac{\epsilon X_0}{\tau M_{Pl}}
\]
It is straightforward to check that all slow-roll conditions are
satisfied provided that the parameters obey
\[
    \frac{M_{Pl}^4 \tau^2 \zeta^2}{\epsilon^4} \ll 1
\]
and the value of the field $X$ satisfies
\bea
    X_0^2 &\ll& M_{Pl}^2  \frac{M_{Pl}^4 \tau^2 \zeta^2}{\epsilon^4}
\nonumber \\
    X_0^2 &\ll& \frac{\epsilon^2}{\zeta^2}
\nonumber
\eea
Clearly, such a choice can indeed be made.

\section{Perturbations about slowly rolling background}
\label{sec:Pert}

Let us consider  perturbations of the fields $B_\mu$
and $\Phi$ in the slow-roll regime. Assuming that all 
masses are small compared to $M_{Pl}$, we neglect mixing
with perturbations of gravitational field\footnote{Mixing 
with perturbations
of the gravitational field in Minkowski background
leads to new effects
at even lower momentum scales
as compared to those considered here~\cite{Ben, MLVR}. These
scales are well below the Hubble scale in our context,
so they are not of interest here.}. To simplify formulas,
we take $\alpha$ and $\beta$ independent of $X$; our main conclusions
remain valid in the general case.
 
Before proceeding, let us make a few comments 
on the relations between the mass
terms,
the value of $\epsilon$ and the Hubble parameter.
First, we note that eq.~(\ref{negVX}), i.e., the 
negative sign of
$XV_X$, ensures that the mass term for vector perturbations
$b_i$, eq.~(\ref{mbi}), is  positive.
Second, from eqs.~(\ref{co1}) and (\ref{co2}) 
and the slow-roll equation of
motion (\ref{eqroll2}) it follows that
\be
   \epsilon^2 \gg 
      m_1^2   \gg \alpha H^2
\label{massrange1}
\ee
where, as before, $m_1^2 = -\frac{1}{ X} U_X$.
Likewise,  eq.~(\ref{Roll3}) implies that 
$m^2_\varphi \equiv W_{\phi \phi}$ is small. 
Assuming that
$U_{XX} \sim X^{-1}U_X$, we thus come to consider the case
studied in Section~\ref{sec:Minkowski} 
in Minkowski background. 

Due to (\ref{largeepsilon})
and (\ref{massrange1}), the Hubble parameter is
small compared to $\epsilon/\sqrt{\alpha}$, 
$\epsilon/\sqrt{\kappa}$ and $m_1/\sqrt{\alpha}$. Therefore,
the expansion of the Universe has no effect on transverse modes
which have the gap in the spectrum (\ref{otrans})
as well as on 
the two modes with the gaps (\ref{obi}) and (\ref{oeps}).
For given conformal momentum $k$, one mode becomes tachyonic
as the physical momentum redshifts to $p_c$, which is given by
(\ref{maxmom}), and which is much larger than $H$. As the physical
momentum redshifts further, this mode enters super-horizon regime.
Thus, we  have to understand the behavior of this mode in that 
regime\footnote{In the case $m_0^2 = -m_1^2$ 
considered in the end of section \ref{sec:Minkowski}, there are no
tachyonic modes at $p^2 < m_1^2 /\kappa$. The analysis below applies to
the low-momentum ghost mode, which obeys
$\omega^2 = p^2$ in Minkowski space-time.}.

It is convenient to work with  conformal metric,
\[
  ds^2 = a^2(\eta) (d\eta^2 - d\bf{x}^2)
\]
In this metric, the fields with perturbations are
\[
   B_0 = aX + b_0 \; , \;\;\;
   B_i = b_i  \; , \;\;\;
   \Phi = \phi + \frac{1}{a} \chi
\]
where $X$ and $\phi$ are background fields.
Equations for longitudinal $b_i = \frac{k_i}{k} b_L$,
$b_0$ and $\chi$ in conformal time  
are\footnote{We do not write here the terms proportional to 
$X^\prime$. It can be shown that in the slow-roll regime,
these terms are subdominant for
both  super-horizon and sub-horizon modes. We also neglect
the time dependence of masses for the same reason.}
\bea
- \kappa \left( b_0^{\prime \prime}  -\frac{a^{\prime \prime}}{a} b_0
+ k^2 b_0 \right)  
- 3 \alpha \frac{a^{\prime 2}}{a^2}  b_0 - a^2 m_0^2 b_0
+ 
2ik\alpha \frac{a^\prime}{a} b_L + 
\epsilon a \left( \chi^\prime - \frac{a^\prime}{a} \chi \right)
= 0
\label{b0FRW} \\
-\alpha \left( b_L^{\prime \prime} -\frac{a^{\prime \prime}}{a} b_L
- \frac{a^{\prime 2}}{a^2} b_L
+ k^2 b_L \right)  
 - a^2 m_1^2 b_L
- 2ik \alpha \frac{a^\prime}{a} b_0 -
i \epsilon k  a \chi 
= 0
\label{bLFRW} \\
- \chi^{\prime \prime} + \frac{a^{\prime \prime}}{a} \chi - k^2 \chi 
- a^2 m_{\varphi}^2 \chi - \epsilon a \left( b_0^{\prime}
+ 2 \frac{a^{\prime}}{a} b_0 \right) + i \epsilon k a b_L
= 0
\label{chiFRW}
\eea
where $k$ is conformal momentum and prime denotes $\d / \d \eta$.
We are interested in slow modes whose physical momenta are small,
\be
k/a \ll m_{0,1} / \sqrt{\alpha}
\label{momrange1}
\ee
 and time-derivatives are small compared
to $m_{0,1} / \sqrt{\alpha}$. Recalling that $m^2_{0,1} \gg \alpha H^2$,
we neglect the first two terms in eq.~(\ref{b0FRW}) and the first
term in eq.~(\ref{bLFRW}).
Then these two equations become algebraic equations for $b_0$ and $b_L$,
giving
\bea
   b_0 &=& \frac{\epsilon}{m_0^2}
   \left[ \frac{1}{a} \chi^\prime - H \chi -
    \frac{2\alpha H}{m_1^2}   
\left( \frac{k}{a} \right)^2 \chi \right]
\label{b0chi} \\
    b_L &=& \frac{\epsilon}{m_1^2}
   \left[ -i\frac{k}{a} \chi 
    + \frac{2i\alpha H}{m_0^2}   
\frac{k}{a}  \left(\frac{1}{a}\chi^\prime - H \chi
\right) \right]
\label{bLchi}
\eea
where $H= a^\prime / a^2$ is the Hubble parameter, and we made use
of (\ref{massrange1}) and (\ref{momrange1}). Plugging these expressions
into eq.~(\ref{chiFRW}),
we obtain the equation for $\chi$,
\[
- \chi^{\prime \prime} + \frac{a^{\prime \prime}}{a} \chi   -k^2 \chi
+ \frac{\epsilon^2}{m_0^2}
\left[ - \chi^{\prime \prime} +  \frac{a^{\prime \prime}}{a} \chi
+ \frac{m_0^2}{m_1^2}k^2 \chi - a^2 \frac{m_0^2 m_\varphi^2}{\epsilon^2}\chi 
+2\alpha  \frac{k^2}{m_1^2} \left( H^2 -\frac{1}{a}H^\prime \right)
 \chi \right] = 0
\]
Recalling again the relations (\ref{massrange1}), 
we obtain finally the equation for the soft mode, 
\be
   - \chi^{\prime \prime} +  \frac{a^{\prime \prime}}{a} \chi
+ \frac{m_0^2}{m_1^2}k^2 \chi  - a^2 \frac{m_0^2 m_\varphi^2}{\epsilon^2}\chi
=0
\label{finito}
\ee
We see that for $m_\varphi=0$,
the field $\chi$ obeys the
equation for the scalar field with tachyonic dispersion relation
(\ref{omegatach}),
but now in the expanding Universe. For non-vanishing
$m_\varphi$, the relation (\ref{Roll3}) implies that the term
with $m_\varphi^2$ is small compared to the second term (the latter
is of order $H^2 a^2\chi$). 
Thus, the solutions of eq.~(\ref{finito})
have tachyonic behavior at $k/a \gg H$ with the dispersion relation
(\ref{omegatach}).

On the other hand, in super-horizon regime $k/a \ll H$, there are usual
``constant'' mode $\chi \propto a$ and decaying mode (for time-independent
$H$ the latter is $\chi \propto a^{-2}$). More precisely, in de~Sitter
space-time with
\[
   a = -\frac{1}{H \eta}
\]
the ``constant'' mode is
\[
    \chi = -\frac{c}{H \eta^{1-\delta}}
\]
where $c$ is the small amplitude and
\[
   \delta = \frac{m_\varphi^2 m_0^2}{3\epsilon^2 H^2} \ll 1
\]
The perturbation of the field $\Phi$ is, therefore, almost
time-independent,
\be
  \varphi \equiv \frac{1}{a} \chi = c \eta^\delta
\label{slowphi}
\ee
It follows from (\ref{b0chi})
that for the ``constant'' mode one has 
\[
b_0 = \frac{\epsilon H}{m_0^2}~ \delta~ \chi
\]
This means that the 
physical temporal component of the vector field perturbation
also slowly varies in time,
\be
    \frac{b_0}{a} \sim c \frac{m_\varphi^2}{\epsilon H} \eta^\delta
\label{slowB}
\ee
On the 
other hand, it follows from (\ref{bLchi}) that for the ``constant''
mode, $b_L$ does not depend on time, so the physical spatial
component of the vector field, $B_i /a$, decays as $a^{-1}$.

We conclude that in the superhorison regime, the tachyonic mode is no
longer dangerous, as it gets almost frozen in. This mode
has finite time to develop in the expanding Universe.

\section{Conclusions}
\label{sec:Conclusions}

Needless to say, our model needs a lot of fine 
tuning to be a viable candidate for describing
inflation or present-time
cosmological evolution. In particular, the ratio 
$\epsilon^2/ (\alpha H^2)$ is to be
large in the slow roll regime, but not exceedingly large. Indeed, from 
the time a  mode becomes tachyonic (this occurs when 
$k/a = \epsilon /\sqrt{\alpha}$) to the time this mode exits the horizon
(at that time $k/a \sim H$), the growh factor 
for this mode is 
\[
     \mbox{exp} \left(N_{growth} \right) \sim \mbox{exp} 
\left( \frac{1}{H}
       \int_H^{\sqrt{\epsilon^2/\alpha}}~\omega(p) \frac{dp}{p}
\right)
\]
The integral here saturates at $p \sim p_c$, and using
(\ref{omegamin}) one estimates
\[
   N_{growth} = \mbox{const} \cdot \sqrt{\frac{\kappa}{\alpha}}
\cdot
\sqrt{\frac{\epsilon^2}{\alpha H^2}}
\]
For the growth factor not to be huge, one either has to fine tune
$\kappa$ to be much smaller than $ \alpha$, 
or fine tune $\epsilon^2 /(\alpha H^2)$ to be not too
 large, or both\footnote{As we mentioned above,
another possibility is to fine tune
$\frac{m_1^2}{\kappa}=\frac{\epsilon^2 - m_1^2}{\alpha}$.
In that case the tachyon does not exist at all.}. 
Crudely speaking, energy density in inhomogeneous
modes makes a small fraction of the background energy density only if
\[
   \frac{\sqrt{\kappa}}{\alpha} \frac{\epsilon}{H} \ll \log 
\frac{M_{Pl}}{\epsilon}
\]
It would be interesting to see whether
this and numerous other fine tunings can occur automatically. 
  
The model presented here is almost certainly not the most appealing
UV-safe phantom theory. Our attitude was rather to show that such a 
theory is possible at all. We therefore did not attempt to study 
a number of
important issues: whether the slow roll regime can be a cosmological
attractor, whether in inflationary context there is a natural way to
exit from this regime, whether in the context of the recent Universe
the tachyon instability at long
wavelengths may have interesting consequences, etc. 
Even before addressing 
these issues it would be desirable to understand what properties of
this  model are generic, and what properties are
model-dependent. In particular, it would be interesting to see whether
tachyons, which exist in our model in a fairly wide
range of spatial momenta
above $H$, may be avoided, leaving behind less dangerous IR ghosts.

The author is indebted to D.~Gorbunov, D.~Levkov, M.~Libanov, S.~Sibiryakov
and P.~Tinyakov for useful discussions, and to S.~Dubovsky for helpful
correspondence. This work is
supported in part by RFBR grant 05-02-17363-a.

\end{document}